\documentclass[letterpaper, 10 pt, conference]{ieeeconf}  

\IEEEoverridecommandlockouts                              
\usepackage[utf8]{inputenc}
\usepackage[T1]{fontenc}
\usepackage{cite}
\usepackage{amsmath,amssymb,amsfonts}
\usepackage{algorithmic}
\usepackage{graphicx}
\usepackage{textcomp}

\usepackage{booktabs} 
\usepackage{graphicx} 
\usepackage{multirow} 
\usepackage{threeparttable}  
\usepackage{makecell}

\title{\LARGE \bf
Learning High-Order Relationships with Hypergraph Attention-based Spatio-Temporal Aggregation for Brain Disease Analysis}

\author{Wenqi Hu$^{1}$, Xuerui Su$^{1}$, Guanliang Li$^{1}$, Yidi Pan$^{1}$ and Aijing Lin$^{1}$
\thanks{$^{1}$Wenqi Hu, Xuerui Su, Guanliang Li, Yidi Pan, and Aijing Lin are with the School of Mathematics and Statistics, Beijing Jiaotong University, Beijing 100044, China (e-mail: 23121721@bjtu.edu.cn).
       }%
}

\begin{document}

\maketitle
\thispagestyle{empty}
\pagestyle{empty}

\begin{abstract}
  Traditional functional connectivity based on functional magnetic resonance imaging (fMRI) can only capture pairwise interactions between brain regions. Hypergraphs, which reveal high-order relationships among multiple brain regions, have been widely used for disease analysis. However, existing methods often rely on predefined hypergraph structures, limiting their ability to model complex patterns. Moreover, temporal information, an essential component of brain high-order relationships, is frequently overlooked.
  To address these limitations, we propose a novel framework that jointly learns informative and sparse high-order brain structures along with their temporal dynamics. Inspired by the information bottleneck principle, we introduce an objective that maximizes information and minimizes redundancy, aiming to retain disease-relevant high-order features while suppressing irrelevant signals. Our model comprises a multi-hyperedge binary mask module for hypergraph structure learning, a hypergraph self-attention aggregation module that captures spatial features through adaptive attention across nodes and hyperedges, and a spatio-temporal low-dimensional network for extracting discriminative spatio-temporal representations for disease classification.
  Experiments on benchmark fMRI datasets demonstrate that our method outperforms the state-of-the-art approaches and successfully identifies meaningful high-order brain interactions. These findings provide new insights into brain network modeling and the study of neuropsychiatric disorders.
    
  \end{abstract}

  \section{Introduction}
  \label{sec:introduction}
  Functional magnetic resonance imaging (fMRI) has been widely used as a non-invasive tool for measuring brain activity and exploring intrinsic patterns of neural organization \cite{heeger2002does}. Functional connectivity (FC), derived from fMRI, captures neural interaction patterns and disorder-related changes by measuring pairwise statistical dependencies between distinct brain regions \cite{cai2017estimation}. Recent studies show that most brain activities involve interactions among multiple regions \cite{semedo2019cortical}. These high-order relationships cannot always be decomposed into pairwise connections, allowing them to capture information that pairwise measures cannot reach \cite{bick2023higher}. To understand high-order interactions among brain regions, some studies model multi-region relationships using hyperedges, but they often fail to extract meaningful patterns effectively \cite{zhu2018dynamic,xiao2019multi}. This highlights the need for more expressive models to learn interpretable high-order brain structures.

  Beyond brain connectivity, the temporal dynamics of fMRI signals are closely linked to cognitive states and the progression of neurological disorders \cite{kong2024multi,zhu2024temporal}, making temporal information a critical component of brain network analysis. However, most current studies on brain high-order relationships focus only on static hypergraph structures \cite{zhang2022multi,wang2023dynamic}, which may neglect complex and transient spatio-temporal dependencies. To comprehensively capture rich high-order information, it is necessary to develop models that simultaneously incorporate both spatial and temporal dimensions of brain high-order relationships within a unified framework.
  
\begin{figure}[htbp]
      \centering
      \includegraphics[width=\columnwidth]{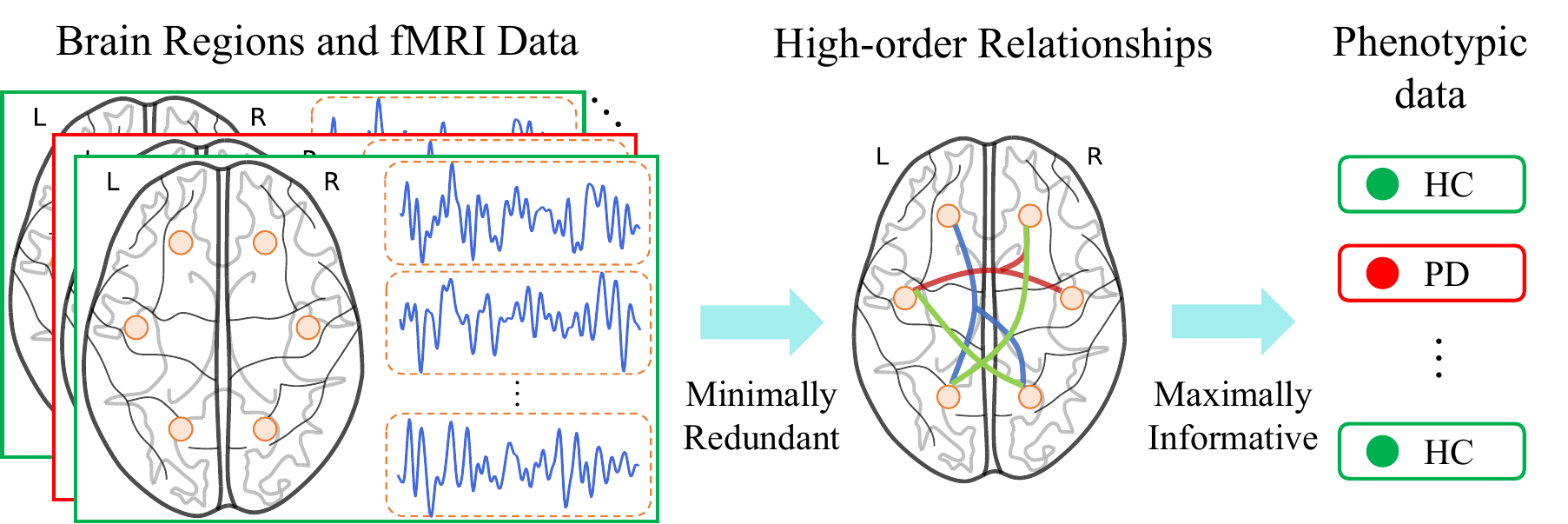}
      \caption{Illustration of high-order relationships with maximum information and minimum redundancy. HC denotes health controls, PD denotes patients with disease.}
      \label{fig:MIMR}
  \end{figure}

  To identify high-order relationships among brain regions that contribute to neurological disorder diagnosis and to capture their temporal dynamics from fMRI signals, we propose a novel \textbf{H}ypergraph \textbf{A}ttention-based \textbf{S}patio-\textbf{T}emporal
  \textbf{A}ggregation framework (HA-STA) that learns the most informative and least redundant high-order structures in the brain. 
  Inspired by the information bottleneck principle \cite{kim2021drop}, we design a new learning objective: Maximize Information and Minimize Redundancy (MIMR), as illustrated in Fig. \ref{fig:MIMR}. This objective aims to preserve high-order relationships that are most predictive of disease while reducing the influence of irrelevant information, enabling the model to discover interpretable hypergraph structures. In addition, we introduce a hyperedge sparsity constraint to reduce topological redundancy caused by irrelevant nodes.
  
  At the model level, we first design multi-hyperedge binary masking to identify the hyperedge structures of brain regions. Then, we propose a \textbf{H}ypergraph \textbf{S}elf-\textbf{A}ttention \textbf{A}ggregation (HSAA) module to extract spatial features from the hypergraph. This module includes hyperedge-level and node-level self-attention mechanisms to learn adaptive attention coefficients for feature aggregation across nodes and hyperedges. To more effectively extract spatio-temporal features and obtain more discriminative low-dimensional representations, we further design a \textbf{S}patial-\textbf{T}emporal \textbf{L}ow-dimensional \textbf{Net}work (ST-LNet) to process the output of HSAA and perform disease classification.
  
  Experimental results demonstrate that our model consistently outperforms the state-of-the-art approaches across multiple benchmark datasets. It also shows great potential in advancing brain network analysis and its application in neuropsychiatric disorder research. The main contributions of this work are summarized as follows:
  \begin{enumerate}
    \item[1)] We improve the traditional information bottleneck framework by embedding hypergraph structure as a prior. Through a hierarchical sparsity constraint, we explicitly incorporate the topological structure of the hypergraph into the information compression process, addressing the problem of hyperedge redundancy.
  
    \item[2)] We design the ST-LNet module, which enables more accurate spatiotemporal feature extraction and yields more informative low-dimensional representations, thereby improving disease prediction performance.
  
    \item[3)] The learnable hyperedges reveal brain regions and high-order interactions most relevant to disease diagnosis. The results show that high-order relationships frequently exist among large groups of nodes, whereas fixed hyperedge construction rules often lead to insufficient feature learning. Moreover, our method identifies discriminative and physiologically meaningful nodes and hyperedges, which are consistent with findings from prior neuroscience research, offering new perspectives for future analysis of high-order brain networks.
  \end{enumerate}
  
  \section{Related Work}
  
  \subsection{Hypergraph Construction in Brain}
  
  Hypergraph construction is a critical step in hypergraph-based brain network modeling. Existing methods are generally divided into two categories, i.e., distance-based methods \cite{huang2009video,gao20123}, representation-based methods \cite{liu2016elastic,liu2017view}. 
  Distance-based methods utilize distances in the feature space to mine the relationships between nodes. Typically, two strategies are employed to construct hyperedges: nearest neighbor search and clustering. For instance, Ji et al. \cite{ji2022fc} combined k-nearest neighbor similarity with k-means clustering to identify the nearest direct neighbors and similar nodes around shared centroids.
  On the other hand, representation-based methods define hyperedges based on feature reconstruction. For example, Wee et al. \cite{wee2014group} constructed functional connectivity using group Lasso with an $l_{2,1}$-norm regularizer to classify patients with mild cognitive impairment from normal controls. Jie et al. \cite{jie2016hyper} adopted a $l_{0}$-norm sparse regression algorithm to construct a hyper-connectivity network for each subject.
  However, the scale of the hyperedges of these methods is limited by the number of nodes, which may affect the performance of hypergraph learning.
  Moreover, hyperedges constructed by these methods tend to vary across subjects, resulting in inconsistent representations.
  
  In contrast to these methods, we propose a deep learning-based framework for learning consistent hypergraph structures. The learned hyperedges are theoretically guaranteed to satisfy the MIMR, while reducing topological redundancy.

  \subsection{Brain Disorder Identification Using Hypergraphs and Spatial-Temporal Features}
  
  Many studies have captured spatio-temporal features by constructing dynamic FC networks.
  For example, Jia et al. \cite{jia2020graphsleepnet} proposed an adaptive graph learning mechanism to extract spatio-temporal information by joint spatio-temporal graph-attentive convolutional networks. Yap et al. \cite{yap2024deep} proposed a deep spatio-temporal variational Bayes framework to learn time-varying topologies in dynamic FC networks.
  These methods are effective in capturing the temporal evolution of pairwise connections, but do not focus on high-order relationships.
  In recent, some studies have successfully utilized both temporal and higher-order features in FC networks classification tasks.
  Teng et al. \cite{teng2024constructing} constructed a high-order functional connectivity network with temporal information by fitting state transitions of fMRI time series to capture temporal dependencies and employing a k-nearest neighbor strategy to obtain a hypergraph.
  Liu et al. \cite{liu2023deep} proposed a temporal and spatial weighted hypergraph connectivity network that fuses high-order spatial relationships and temporal hypercorrelations based on a predefined hyperedge weight calculation function.
  These methods considered high-order relationships between brain regions, but neglected the correlations between brain regions and high-order relationships.
  The generative rules and unlearnability of hyperedges lead to a lack of interpretability of the established high-order relationships.
  
  To address the limitations of these methods, based on learning the high-order relationships of MIMR, we introduce hyperedge attention aggregation for extracting correlations between brain regions and high-order relationships, and propose the ST-LNet framework for fusing high-order relationships and temporal information for disease diagnosis.
  
  \section{Method}
  \label{sec1}
\begin{figure*}[htbp]
      \centering
      \includegraphics[width=\textwidth]{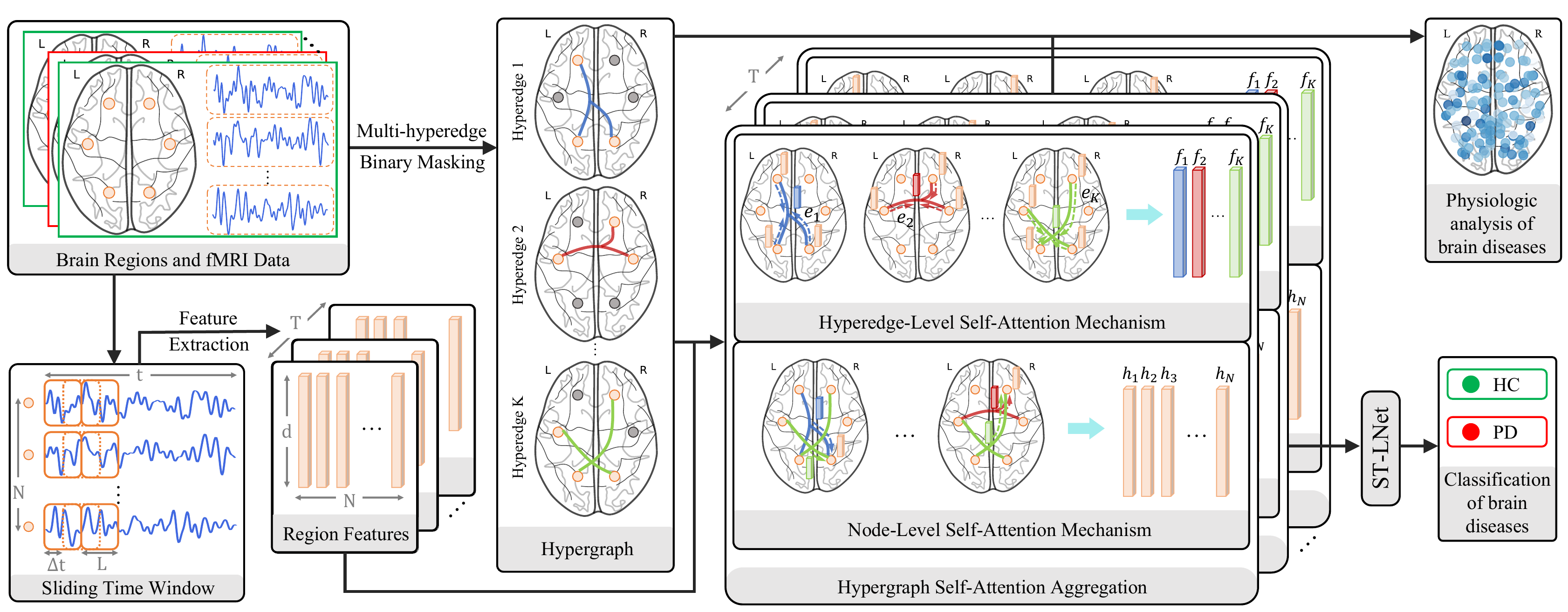}
      \caption{Illustration of the proposed framework. Firstly, region features are extracted based on inter-region correlation. Secondly, Multi-hyperedge Binary Masking is adopted to learn the hyperedges. Then, the correlation information between nodes and hyperedges is fused by HSAA. Finally, ST-LNet is adopted to fuse spatio-temporal information, and the deeply fused features are used for classification and analysis of brain diseases.}
      \label{fig:framework}
  \end{figure*}

  The proposed hypergraph attention-based spatio-temporal aggregation framework is illustrated in Fig. \ref{fig:framework}. It consists of the following steps: 1) Extract region features from dynamic FC networks; 2) Construct hypergraph to represent high-order relationships using multi-hyperedge binary masking; 3) Apply HSAA to integrate relational information between hyperedges and brain regions; 4) Aggregate spatio-temporal features via ST-LNet for brain disease classification. These steps correspond to the subsequent subsections.
  
  \subsection{Dynamic Functional Connectivity Networks}
  To construct dynamic FC networks from fMRI data, we adopt a sliding window approach. The fMRI time series of each brain region is segmented into multiple overlapping fragments, where the window length is denoted by $L$ and the step size by $\Delta t$. For each time window, a correlation matrix is computed to capture the FC patterns among brain regions.
  
  Specifically, at time window $t$, the correlation matrix is represented as
  \begin{equation}
    X_{t} = [X^{1}_{t}, X^{2}_{t}, \dots, X^{N}_{t}]^{T} \in \mathbb{R}^{N \times d},
  \end{equation}
  where $N$ is the total number of brain regions and $X_{t,i} \in \mathbb{R}^{d}$ is the column vector corresponding to brain region $i$, representing its FC strengths with all other regions. We use $X_{t,i}$ as the feature vector of brain region $i$ at time window $t$.
  
  \subsection{Hypergraph Construction via Multi-hyperedge Binary Masking}
  
  Traditional adjacency matrices can only capture pairwise relationships between nodes, whereas hypergraph representations are capable of modeling high-order relationships involving more than two nodes, thereby providing a more comprehensive characterization of brain structures. In our framework, the brain regions defined by a predefined atlas serve as the nodes of the constructed hypergraph. To construct each hyperedge, we assign it a binary mask vector responsible for selecting the nodes belonging to the corresponding hyperedge.
  
  Suppose $K$ denote the predefined number of hyperedges. 
  Formally, for the $k$-th hyperedge $m^k \in \{0,1\}^N$, each element in the vector corresponds to a brain region:
  \begin{equation}
    \label{1}
      m^k = [\mathbf{1}(p^k_{\theta,1}), \mathbf{1}(p^k_{\theta,2}), \dots, \mathbf{1}(p^k_{\theta,N})]^T \in \{0,1\}^{N},
  \end{equation}
  where $p^k_{\theta,i} \in [0,1]$ are learnable probabilities, and the indicator function $\mathbf{1} : [0,1] \to \{0,1\}$ is defined as:
  \begin{equation}
      \mathbf{1}(x) =
      \begin{cases} 
          1, & \text{if } x > 0.5 \\ 
          0, & \text{if } x \leq 0.5 
      \end{cases}
  \end{equation}
  Since the indicator function is non-differentiable, we approximate the gradient using the stop-gradient technique \cite{van2017neural} to enable backpropagation through the masking operation motivated by the technique of the categorical reparameterization with Gumbel-Softmax \cite{jang2016categorical}.
  
  In the binary mask vector $m^k$, $0$ signifies that the corresponding node is excluded from the hyperedge, and $1$ indicates that the node is included. All included nodes are considered to form a hyperedge together.
  At time window $t$, we denote the masked representation of $X_{t}$ for the $k$-th hyperedge as:
  \begin{equation}
    \boldsymbol{e}_{k} = m^k \odot X_{t} = [m^k_1 X_t^1, m^k_2 X_t^2, \dots, m^k_N X_t^N],
  \end{equation}
  where $\odot$ represents element-wise multiplication broadcasted across dimensions. $m^k_j$ denotes the $j$-th element of the mask vector $m^k$. The masked representation $\boldsymbol{e}_k$ responds to the features of brain regions contained in the $k$-th hyperedge.
  
  \subsection{Hypergraph Self-Attention Aggregation}
  In order to efficiently capture the associations between nodes and hyperedges and highlight the importance of different nodes on hyperedges, inspired by graph attention mechanisms \cite{ji2022fc,zhu2024dynamical}, we apply self-attention mechanisms on hypergraphs to learn the hidden representations of each node and hyperedge, called hypergraph self-attention aggregation. This method  consists of two primary attention mechanisms: hyperedge-level self-attention mechanism and node-level self-attention mechanism, both of which leverage learnable attention coefficients to adaptively aggregate node and hyperedge features.
  
  \subsubsection{Hyperedge-Level Self-Attention Mechanism}
  Since not all brain regions are equally relevant to group-level activity, we introduce a self-attention mechanism to emphasize those nodes that are most critical for understanding each hyperedge. This mechanism allows the model to aggregate node features into hyperedge-level representations based on their importance.
  
  Specifically, with brain region nodes $\mathcal{V} = \{v_i\}$ and hyperedges $\mathcal{E} = \{e_j\}$, we denote $h_s^{l-1}$ as the feature representation of node $v_s$ at the $(l-1)$-th layer. To obtain the representation of hyperedge $e_j$, we aggregate the features of its associated nodes using attention coefficients:
  \begin{equation}
  f_j^l = \sigma \left( \sum_{v_s \in e_j} \alpha_{js} W_1 h_s^{l-1} \right),
  \end{equation}
  where $ \sigma(\cdot) $ is a non-linear activation function and $ W_1 $ is a trainable weight matrix. The attention coefficient $ \alpha_{js} $ determines the importance of node $ v_s $ to hyperedge $ e_j $, which is computed as:
  \begin{equation}
  \alpha_{js} = \frac{\exp(a_1^\top u_s)}{\sum_{v_p \in e_j} \exp(a_1^\top u_p)}, 
  \end{equation}
  where
  \begin{equation}
  u_s = \text{LeakyReLU} (W_1 h_s^{l-1}),
  \end{equation}
  and $ a_1^\top $ is a learnable weight vector. 
  
  \subsubsection{Node-Level Self-Attention Mechanism}
  To incorporate information from hyperedges and enhance node embeddings, we introduce a node-level self-attention mechanism that highlights the most informative hyperedges for updating each node representation. The update equation for node features is given by:
  \begin{equation}
  h_i^l = \sigma \left( \sum_{e_j \in \varepsilon_i} \beta_{ij} W_2 f_j^l \right),
  \end{equation}
  where $ W_2 $ is a trainable weight matrix, and $ \varepsilon_i $ denotes the set of hyperedges connected to node $ v_i $. The attention coefficient $ \beta_{ij} $ that determines the importance of hyperedge $ e_j $ to node $ v_i $ is computed as:
  \begin{equation}
  \beta_{ij} = \frac{\exp(a_2^\top z_j)}{\sum_{e_p \in \varepsilon_i} \exp(a_2^\top z_p)},
  \end{equation}
  where
  \begin{equation}
  z_j = \text{LeakyReLU} \left( [W_2 f_j^l \parallel W_1 h_i^{l-1}] \right),
  \end{equation}
  and $ a_2^\top $ is another learnable attention vector, and $ \parallel $ denotes concatenation. The final node representation $ h_i^l $ is obtained by aggregating hyperedge information through these learned attention weights.
  

  \subsection{Spatio-Temporal Low-dimensional Network}
\begin{figure*}[htbp]
    \centering
    \includegraphics[width=\textwidth]{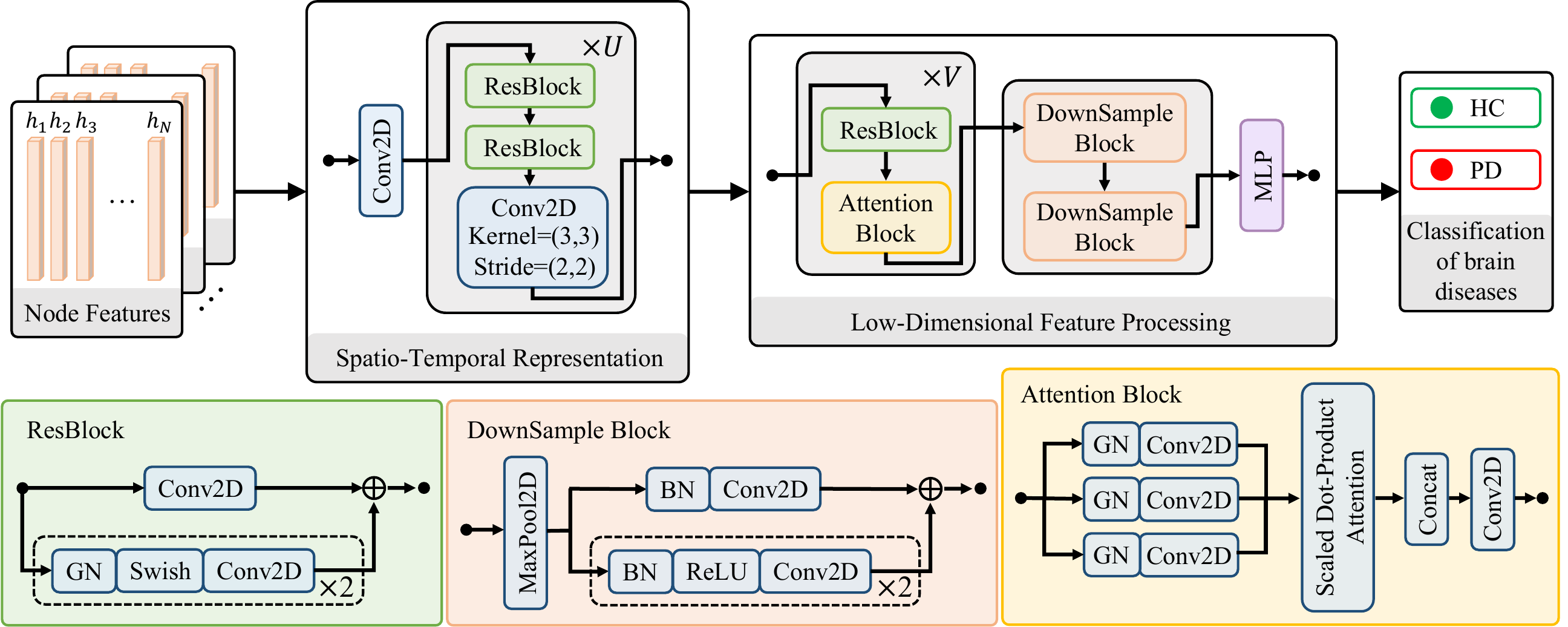}
    \caption{Illustration of ST-LNet. BN denotes batch normalization and GN denotes group normalization.}
    \label{fig:ST-LNet}
  \end{figure*}

  In order to better mine spatio-temporal feature information and obtain more scientific and reasonable low-dimensional feature representations, so as to improve the performance of disease prediction, we designed Spatio-Temporal Low-dimensional Network, which is composed of two parts: Spatio-Temporal Representation module and Low-Dimensional Feature Processing module. The details of the feedforward process and related modules of ST-LNet are shown in Fig. \ref{fig:ST-LNet}.
  
  Specifically, the Spatio-Temporal Representation module comprises a resolution-preserving Conv2D head layer followed by $U$ dimensionality-reduction submodules, each containing two ResBlocks and a downsampling Conv2D layer. The input spatio-temporal feature map ${h}_{1:N}\in\mathbb{R}^{T\times N \times D}$ (where $D$ is the feature dimension) first passes through the head layer for preliminary encoding, and then sequentially through the $U$ submodules to achieve gradual compression and extraction of spatio-temporal features:
  \begin{equation}
    \begin{aligned}
    \mathbf{H}^{(0)} =& \mathrm{Conv2D}_{\mathrm{head}}(h_{1:N}), \\
    \widetilde{{\mathbf{H}}}^{(u)} =& \mathrm{ResBlock}_u^2\left(\mathrm{ResBlock}_u^1\bigl(\mathbf{H}^{(0)}\bigr)\right), \\
    {{\mathbf{H}}}^{(u)} =& \mathrm{Conv2D}^{\downarrow}_u\bigl(\widetilde{{\mathbf{H}}}^{(u)}\bigr),
    \end{aligned}
    \end{equation}
  where $u=1,\dots,U$.
  
  Secondly, the Low-Dimensional Feature Processing module consists of $V$ attentional low-dimensional representation submodules, two DownSample Blocks, and a MLP output layer. Each attentional low-dimensional representation submodule integrates a ResBlock and an Attention Block to further enhance the expressiveness of low-dimensional features. Formally, given $\mathbf{H}^{(U)}$, the $v$-th attentional submodule computes:
  \begin{equation}
    \widetilde{\mathbf{F}}^{(v)} = \mathrm{ResBlock}_v\bigl({{\mathbf{H}}}^{(U)} \bigr), 
  \quad
  \mathbf{F}^{(v)} = \mathrm{Attn}_v\bigl(\widetilde{\mathbf{F}}^{(v)}\bigr),
  \end{equation}
  where $v=1,\dots,V$, and the aggregated feature is then downsampled twice:
  \begin{equation}
    \hat{\mathbf{y}} = \mathrm{MLP}\left(\mathrm{Flatten}\left(\mathrm{DSB}_2\left(\mathrm{DSB}_1\bigl(\mathbf{F}^{(V)}\bigr)\right)\right)\right),
  \end{equation}
  where each DownSample Block (DSB) comprises MaxPool2D, BatchNorm, a resolution-preserving Conv2D, and ReLU activation function, and $\hat{\mathbf{y}}$ serves as the logits for binary classification.
  
  \subsection{Variational Information Bottleneck Guided Hypergraph Structure Learning}
  
  Motivated by the graph information bottleneck principle \cite{sun2022graph,qiu2023learning}, we introduce variational information bottleneck to learn the hypergraph structure.
  The hypergraph structure learning problem considered in this paper can be formulated as, given brain region feature matrix $X$ and label $Y$, generating an optimized hypergraph $G$ and its corresponding hypergraph representation for accurate prediction of disease outcomes.
  We consider $X$, $Y$ and $G$ are random variables in the Markovian chain $X \leftrightarrow Y \leftrightarrow G$. According to the MIMR objective, the optimization problem can be formulated as follows:
  \begin{equation}
    \label{3}
    G = \arg \min_{G} -I(G ; Y)+\lambda I(G ; X),
  \end{equation}
  where $I(\cdot ;\cdot )$ denotes the mutual information between random variables, and $\lambda$ is a coefficient controlling the trade-off between informativeness and redundancy.

  \subsubsection{Lower bound of informativeness}
  By introducing the variational approximation, there is
  \begin{equation}
    \begin{aligned}
    I(G ; Y)= & \mathbb{H}[Y]-\mathbb{H}[Y | G] \\
    = & \mathbb{H}[Y]+\mathbb{E}_{p(Y, G)}[\log p(Y | G)] \\
    = & \mathbb{H}[Y]+\mathbb{E}_{p(Y, G)}\left[\log q_\phi(Y |G)\right] \\
    & +\mathbb{E}_{p(G)}\left[\mathcal{D}_{KL}\left(p(Y | G) \parallel q_\phi(Y | G)\right)\right] \\
    \geq & \mathbb{H}[Y]+\mathbb{E}_{p(Y, G)}\left[\log q_\phi(Y | G)\right],
    \end{aligned}
  \end{equation}
  where $\mathbb{H}[\cdot]$ denotes the entropy and $\mathcal{D}_{KL}(\cdot \parallel \cdot)$ represents the Kullback-Leibler (KL) divergence. The distribution $q_{\phi}(Y|G)$ is a variational approximation of the true posterior $p(Y|G)$. Since the entropy of $Y$ does not involve any learnable components, the optimization primarily focuses on the second term $\mathbb{E}_{p(Y,G)}[\log q(Y|G)]$. In practice, we model $q_{\phi}$ as the composition of ST-LNet and HSAA with parameters $\phi$. 
  
  \subsubsection{Upper bound of redundancy}
  According to the conclusions in \cite{kim2021drop}, we have:
  \begin{equation}
      \label{2}
      \begin{aligned}
      I(G ; X) \leq \sum_{k=1}^K I\left(\boldsymbol{e}^k ; X\right) & \leq \sum_{k=1}^K \sum_{i=1}^N I\left(\boldsymbol{e}_i^k ; X_i\right) \\
      & =\sum_{k=1}^K \sum_{i=1}^N \mathbb{H}\left[X_i\right]\left(1-p_{\theta, i}^k\right),
      \end{aligned}
  \end{equation}
  where $\boldsymbol{e}_i^k$ and $X_i$ is the $i$-th row of $\boldsymbol{e}^k$ and $X$ respectively. $p_{\theta, i}^k$ is the mask probability in Eq. \ref{1}. The equality holds if and only if nodes are independent and hyperedges do not overlap. It is important to clarify that optimizing Eq. \ref{2} does not imply a penalty for overlaps.
  
  \subsubsection{Sparse constraints on hyperedges}
  We constraint the mask probability of hyperedges $p_{\theta}^k = (p_{\theta,1}^k, p_{\theta,2}^k,\dots, p_{\theta,N}^k)$ by introducing $l_{1}$ normalization:
  \begin{equation}
      \mathcal{L}_{S}  = \sum_{k=1}^K \| p_{\theta}^k \|_{1}.
  \end{equation}
  By minimizing $\mathcal{L}_{S}$, the model takes into account the sparsity of the hyperedges while learning the MIMR, which helps to retain significant hyperedges and enhances structural interpretability.
  
  Therefore, instead of optimizing the intractable objective Eq. \ref{3}, we optimize its upper bound:
  \small{\begin{equation}
  \begin{aligned}
  \mathcal{L} & =\mathcal{L}_{CE}(F_{\phi}(G),Y)+\lambda \sum_{k=1}^K \sum_{i=1}^N \mathbb{H}\left[X_i\right]\left(1-p_{\theta, i}^k\right)+\gamma \mathcal{L}_{S},
  \end{aligned}
  \end{equation}
  }
  where $\mathcal{L}_{CE}$ denotes the cross-entropy loss, $F_{\phi}(G)$ represents the combined ST-LNet and HSAA modules parameterized by $\phi$ and $\gamma$ is a coefficient that controls the strength of sparsity restriction. 
  
  \section{Experiments and Results}
  
  \subsection{Data Acquisition and Preprocessing}
  
  \subsubsection{Datasets and Preprocessing}
  
  In this study, we utilize two publicly available neuroimaging databases: the ADHD-200 database and the Autism Brain Imaging Data Exchange I (ABIDE-I) database. Both databases provide resting-state functional magnetic resonance imaging (rs-fMRI) data for studying neurodevelopmental disorders.
  
  The ADHD-200 database publicly shares neuroimaging data from 362 children and adolescents diagnosed with attention-deficit/hyperactivity disorder (ADHD) and 585 developmentally healthy controls. 
  For preprocessing, we employ the Configurable Pipeline for the Analysis of Connectomes (C-PAC) with specific preprocessing steps including motion correction, slice timing correction, nuisance signal regression, band-pass filtering, and spatial normalization to the standard MNI152 template. The preprocessed rs-fMRI data are publicly accessible from the Preprocessed Connectomes Project (PCP) \cite{bellec2017neuro}.
  In the ADHD-200 Database, we partition brain space into ROIs via the AAL-116 template. This Deterministic atlas is the result of an automated anatomical parcellation of the spatially normalized single-subject high-resolution T1 volume \cite{tzourio2002automated}.
  
  The ABIDE-I database openly shares neuroimaging data from 539 individuals suffering from autism spectrum disorder (ASD) and 573 healthy controls. 
  We use the same preprocessing steps and acquired preprocessed rs-fMRI data from the publicly available Preprocessing Connectome Project \cite{craddock2013neuro}.
  In the ABIDE-I database, the Craddock 200 (CC200) atlas is adopted to partition brain space. This functional parcellation is derived by normalized cut spectrum clustering of voxel connectivity maps within the gray matter mask \cite{craddock2012whole}.

  \subsubsection{Experimental Setup}
  We selected the functional connectivity computed with Pearson correlation coefficient as the features of the regions, and set the time window size $L = 20$ and the time step $\Delta t = 10$.
  The model is trained using Adam optimizer with an initial learning rate of 0.0001 and a weight decay of 0.001. The batch size is set to 16, and the model is trained for 300 epochs with an early stopping criterion based on validation loss.
  We randomly split the data into training set, validation set and test set with a ratio of 8:1:1.

  To evaluate model performance, we employ accuracy (ACC), sensitivity (SEN), specificity (SPE) as primary metrics.
  Beyond these standard evaluation metrics, we further assessed the importance of individual brain regions and hyperedges. 
  To better characterize the role of each brain region under different fMRI tasks, we studied the frequency with which each region appeared in the identified hyperedges. This frequency can be considered as a measure of region importance, and is formally defined for node $i$ as:
  \begin{equation}
    D_i = \frac{1}{|\mathcal{E}|} \sum_{e \in \mathcal{E}} \mathbb{I}(i \in e)
  \end{equation}
  where $\mathcal{E}$ denotes the set of all hyperedges and $\mathbb{I}(\cdot)$ is the indicator function that equals 1 if node $i$ is a member of hyperedge $e$, and 0 otherwise.
  
  Building upon this, we further defined the hyperedge importance $I_e$ by integrating the learned attention coefficients with the region frequencies:
  \begin{equation}
    I_e = \sum_{i \in e} \left(  \alpha_{e,i} \times D_i \right)
  \end{equation}
  where $\alpha_{i,e}$ denotes the attention coefficient assigned by hyperedge $e$ to node $i$, and the summation aggregates the attention-weighted frequencies of all nodes connected by $e$.
  Regional importance and hyperedge importance offer neurobiologically meaningful interpretations of the learned hypergraph structures.
  
  \subsubsection{Baseline Methods}
  We compare HA-STA model against a comprehensive set of baseline methods, which are categorized into three groups: 1) Traditional Machine Learning: This category includes RFE\_SVM \cite{craddock2009disease} and LASSO \cite{li2019multimodal}, which apply classical feature selection and regression techniques to neuroimaging data; 2) Graph Deep Learning: We consider several recent graph-based models such as BrainGNN \cite{mahmood2021attend}, MDCN \cite{yang2023deep}, BNC-DGHL \cite{ji2022functional}, and FC-HAT \cite{ji2022fc}, which utilize graph neural networks to learn representations from brain connectivity structures; 3) Spatio-Temporal Integration: This group includes STGCN \cite{gadgil2020spatio}, STAGIN \cite{kim2021learning}, ST-HAG \cite{liu2024spatio}, and STIGR \cite{liu2025dynamic}, which integrate both spatial and temporal information for brain disorder classification.

  \subsection{Classification performance}
  \subsubsection{Results on ADHD-200 datasets}
  \begin{table*}[h]
    \centering
    \caption{Classification performance on ADHD-200 and ABIDE-I}
    \begin{tabular}{lcccccccc}
        \toprule
        \multirow{2.5}{*}{\makecell{Type}}  &  \multirow{2.5}{*}{\makecell{Model}}  &\multirow{2.5}{*}{\makecell{Year}}&  \multicolumn{3}{c}{ADHD-200}&  \multicolumn{3}{c}{ABIDE-I} \\
        \cmidrule(lr){4-6}  \cmidrule(lr){7-9}& & &ACC (\%) & SEN (\%) & SPE (\%) & ACC (\%) & SEN (\%) & SPE (\%) \\
        \midrule
        \multirow{2}{*}{\makecell{Traditional Machine Learning}}
        &RFE\_SVM &2009  & 63.8 & 70.1 & 51.5 & 63.9 & 63.0 & 64.8 \\
        &LASSO &2019  & 67.6 & 70.4 & 60.1 & 63.4 & 60.9 & 66.0 \\
        \midrule
        \multirow{4}{*}{\makecell{Graph Deep Learning}}
        &BrainGNN &2021  & 72.9 & 78.9 & 61.3 & 68.7 & 69.8 & 65.8 \\
        &MDCN &2023  & 67.5 & 72.0 & 62.4 & 72.4 & 73.2 & 71.7 \\
        &BNC-DGHL &2022 &  69.1 & 75.0 & 61.4 & 68.9 & 63.5 & \textbf{75.0} \\
        
        &FC-HAT &2022  & 69.2 & \textbf{83.0} & 46.8  & 70.9 & 70.0 & 72.3 \\
        \midrule
        \multirow{4}{*}{\makecell{Spatio-Temporal Integration}}
        &STGCN &2020 &  72.6 & 77.5 & 65.1  &  68.2 & 71.5 & 64.2 \\
        &STAGIN &2021 &  73.2 & 75.0 & 66.0  &  69.5 & 74.6 & 64.1 \\
        &ST-HAG &2024 &  74.8 & 78.2 & 68.5  & 71.9 & 72.1 & 68.8 \\
        &STIGR &2025 &  76.2 & 79.5 & 72.4  & 72.7 & \textbf{76.1} & 68.8 \\
        \midrule
  
        Our Proposed &HA-STA &2025& \textbf{80.8} & \underline{80.7} & \textbf{81.0} & \textbf{73.5} & \underline{74.4} & \underline{72.5} \\
        \bottomrule
    \end{tabular}
    \begin{tablenotes}
      \centering
      \footnotesize  
      \item Bold means best, underline means sub-optimal.
    \end{tablenotes}
    \label{tab:performance}
  \end{table*}
  From Table \ref{tab:performance}, HA-STA achieves the best performance on ADHD-200 datasets in terms of ACC and SPE. Although FC-HAT slightly outperforms in SEN, its lower SPE suggests a bias toward positive predictions. In contrast, HA-STA achieves a more balanced classification, indicating its ability to detect ADHD-related patterns without overfitting to a particular class. This reflects the robustness and practicality of HA-STA in handling real-world diagnostic tasks where both SEN and SPE are important.
  \subsubsection{Results on ABIDE-I datasets}
  As shown in Table \ref{tab:performance}, the proposed HA-STA model achieves the highest ACC on ABIDE-I datasets, surpassing all baselines. While STIGR shows a slightly higher SEN and BNC-DGHL attains the top SPE, HA-STA offers a more balanced performance, with both SEN and SPE remaining competitive. This balance enables more reliable detection of autism spectrum disorder cases while minimizing false positives. The results demonstrate the strong generalization ability of HA-STA in large-scale and heterogeneous neuroimaging scenarios.

  \subsection{Ablation Experiments}
  
  To thoroughly evaluate the contributions of each individual component in our proposed model, we conduct ablation experiments on two benchmark neuroimaging datasets: ADHD-200 and ABIDE-I. These experiments are designed to isolate the effects of key architectural choices, including the binary masking mechanism, HSAA, and ST-LNet. The quantitative results are summarized in Table \ref{tab:ablation} for ADHD-200 and Table \ref{tab:ablation1} for ABIDE-I. Additionally, we present t-SNE visualizations to qualitatively support our findings.
  
  \subsubsection{Effectiveness of the Binary Masking Strategy}
  
  We begin by investigating the role of the learnable mask, which is designed to identify and emphasize discriminative brain regions by selectively preserving or discarding node features. We compare two variants:
  1) Nomask: Do not mask at all, which means all nodes and their features are visible to each attention head during hypergraph construction, regardless of their relevance to the classification task;
  2) Softmask: Remove the binary indicator function and use the learned $p^{k}_{\theta,i}$ directly as soft weights for each node, allowing continuous importance modulation without hard selection.
  
  As shown in both tables, models with Nomask perform the worst in terms of classification ACC. For instance, on the ADHD-200 dataset, this configuration achieves only 60.3\% ACC, while the full model incorporating the binary mask achieves a significantly higher ACC of 80.8\%.
  The Softmask variant yields a modest improvement (65.8\% ACC), indicating that learning a probabilistic importance over regions is better than treating all nodes equally. However, it still underperforms the binary mask, suggesting that discrete selection may provide a clearer inductive bias, effectively suppressing redundant or noisy signals in the brain network. This trend is also consistent on the ABIDE-I dataset. 
  These results validate that the binary masking mechanism plays a critical role in enhancing model robustness and discriminative capability by dynamically selecting task-relevant features during graph construction.
  
  \begin{table}[h]
    \centering
    \caption{Ablation Study on ADHD-200}
    \begin{tabular}{lcccc}
        \toprule
        Model Specification & ACC (\%) & SEN (\%) & SPE (\%)&\\ 
        \midrule
        HA-STA w/ Nomask & 60.3 & 51.6 & 66.7 & \\
        HA-STA w/ Softmask & 65.8 & 51.6 & 76.2 & \\
        HA-STA w/o HSAA & 68.5 & 74.2 & 64.3 & \\
        HA-STA w/o ST-LNet & 67.1 & 74.2 & 61.9 & \\
        HA-STA (\textbf{Our Proposed}) & \textbf{80.8} & \textbf{80.7} & \textbf{81.0} & \\
        \bottomrule
    \end{tabular}
    \label{tab:ablation}
  \end{table}
  
  \begin{table}[h]
    \centering
    \caption{Ablation Study on ABIDE-I}
    \begin{tabular}{lcccc}
        \toprule
        Model Specification & ACC (\%) & SEN (\%) & SPE (\%) & \\
        \midrule
        HA-STA w/ Nomask & 63.9 & 58.1 & 70.0 & \\
        HA-STA w/ Softmask & 61.5 & 62.8 & 60.0 & \\
        HA-STA w/o HSAA & 63.9 & 60.5 & 67.5 & \\
        HA-STA w/o ST-LNet & 62.7 & 62.8 & 62.5 & \\
        HA-STA (\textbf{Our Proposed}) & \textbf{73.5} & \textbf{74.4} & \textbf{72.5} &\\
        \bottomrule
    \end{tabular}
    \label{tab:ablation1}
  \end{table}
  
  \subsubsection{Efficacy of Hypergraph Self-Attention Aggregation}
  
  To assess the efficacy of the HSAA module, we replace it with a simpler aggregation method: mean pooling across node features. This substitution leads to a noticeable decrease in ACC from 80.8\% to 68.5\% on ADHD-200 and from 73.5\% to 63.9\% on ABIDE-I, with similarly significant decreases in SEN and SPE. This indicates that mean pooling fails to capture the complex and high-order interactions among brain regions that HSAA is capable of modeling.
  HSAA allows the model to assign varying levels of importance to different nodes within hyperedges, effectively learning more expressive representations.
  
  \subsubsection{Efficacy of Spatio-Temporal Low-dimensional Network}
  
  We further compare two strategies for aggregating spatio-temporal information across time windows: Gated Recurrent Unit (GRU) approach and our proposed fusion network ST-LNet. While GRU are widely used for modeling temporal dependencies, we observe that replacing ST-LNet with GRU leads to a significant decrease in performance. For example, on ADHD-200, ACC drops from 80.8\% to 67.1\%, and on ABIDE-I, from 73.5\% to 62.7\%. The SEN and SPE decrease similarly on both datasets.
  This suggests that although GRU can capture temporal dependencies, spatial features from each time window need to be combined in a more discriminative manner.
  
\begin{figure}[htbp]
    \centering
    \includegraphics[width=\columnwidth]{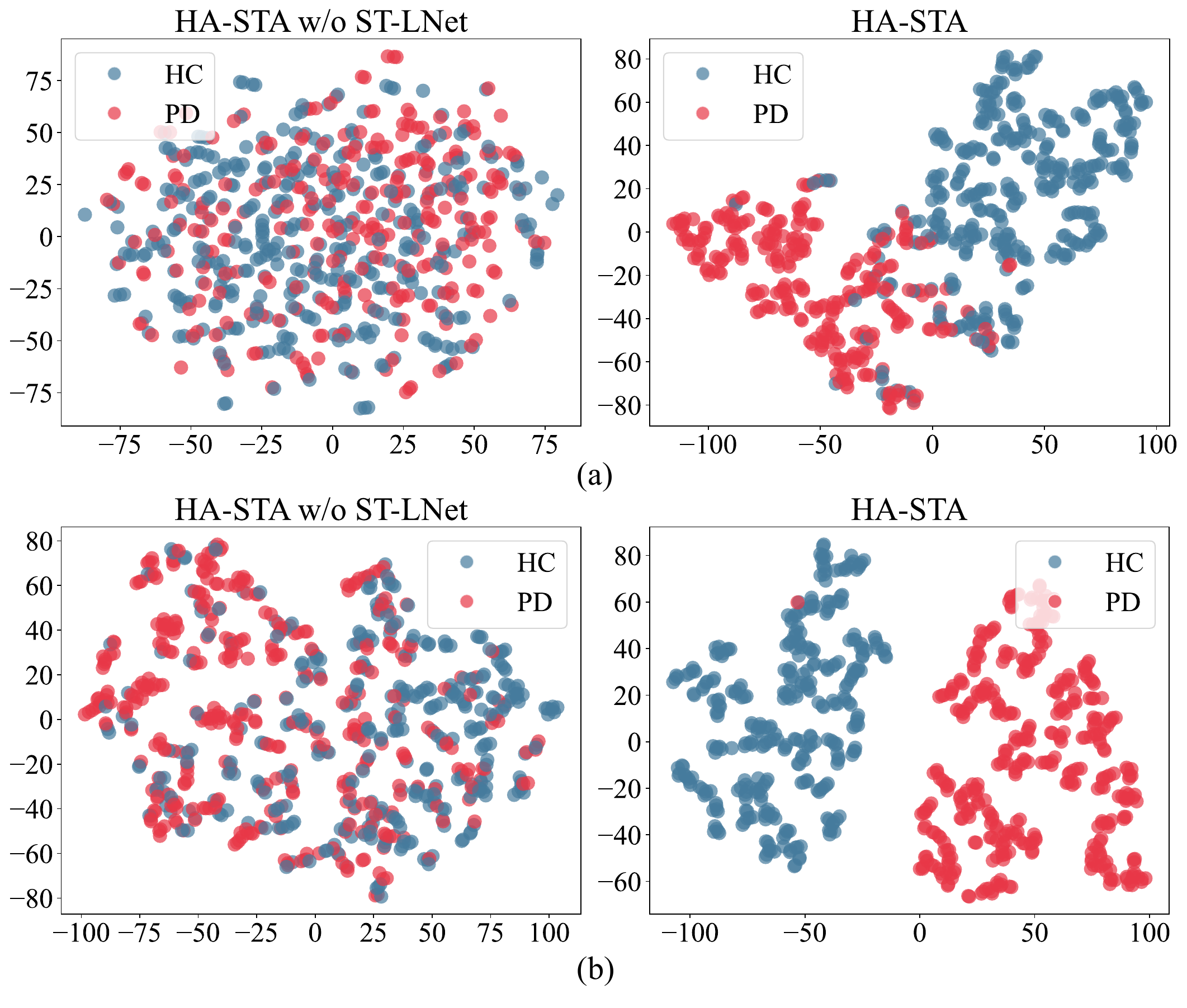}
    \caption{t-SNE visualization of HA-STA and HA-STA w/o ST-LNet on ADHD-200 and ABIDE-I datasets.}
    \label{tsne}
  \end{figure}

  To better understand the quality of the learned representations, we visualize the high-dimensional features using t-distributed stochastic neighbor embedding (t-SNE) \cite{van2008visualizing} for two model variants: HA-STA w/o ST-LNet and HA-STA. 
  The t-SNE plots (Fig. \ref{tsne}) show that our model results in more compact and well-separated clusters corresponding to healthy controls and patients with diseases. In contrast, the GRU-based variant exhibits significant overlap between classes. This highlights the advantage of ST-LNet in extracting discriminative spatio-temporal features, which allows our model to have better classification capabilities.

  \subsection{Parameter Analysis}
  \subsubsection{Effect of regularization hyperparameters}
  \label{regularization_hyperparameters}
  To further investigate the effect of the regularization terms in our objective function, we perform a parameter analysis on the two hyperparameters: $\lambda$ and $\gamma$. $\lambda$ controls the strength of the information bottleneck regularization, which encourages the model to suppress irrelevant information by penalizing uninformative regions selections. $\gamma$ controls the sparsity of the hyperedge structure, ensuring that each hyperedge selectively attends to more informative nodes. 
  \begin{figure}[htbp]
    \centering
    \includegraphics[width=\columnwidth]{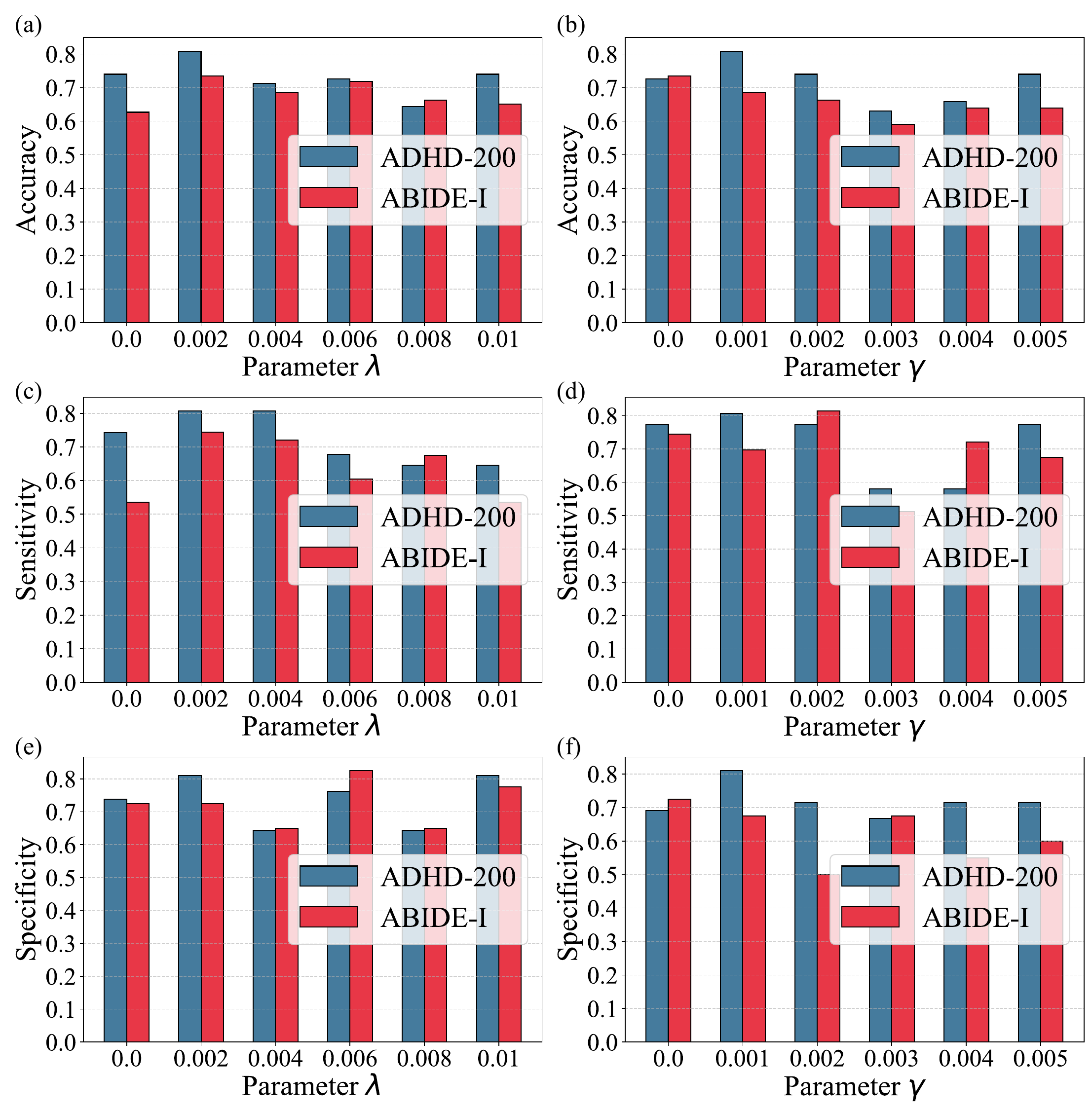}
    \caption{The influence of $\lambda$ and $\gamma$ on HA-STA in terms of ACC, SEN and SPE on ADHD-200 and ABIDE-I datasets.}
    \label{fig:lambda_gamma}
  \end{figure} 
  
  In Fig. \ref{fig:lambda_gamma}(a), the ACC of the ADHD-200 dataset peaks when $\lambda = 0.002$ and drops significantly as $\lambda$ increases, indicating that overly strong regularization impairs the ability of the model to preserve discriminative information. ABIDE-I, on the other hand, exhibits more stable performance across different $\lambda$ values, with a modest peak at $\lambda = 0.002$. In Fig. \ref{fig:lambda_gamma}(b), the effect of $\gamma$ is examined. On ADHD-200, ACC reaches the highest value at $\gamma = 0.001$ and decreases sharply for $\gamma \geq  0.003$, showing that excessive sparsity leads to unstable learning. ABIDE-I performs best when $\gamma = 0$ and gradually declines as $\gamma$ increases, suggesting that sparsity is less beneficial for datasets with more diverse features.
  
  SEN trends in Fig. \ref{fig:lambda_gamma}(c) mirror the ACC results. For ADHD-200, SEN is high at $\lambda = 0.002$ but drops as $\lambda$ increases, reflecting a loss in true positive predictions under strong regularization. ABIDE-I shows stable SEN at lower $\lambda$ values, with minor improvements at intermediate levels. In Fig. \ref{fig:lambda_gamma}(d), SEN on ADHD-200 again peaks at $\gamma = 0.001$, while ABIDE-I achieves the highest SEN at $\gamma = 0.002$, although this comes at the cost of a dramatic decrease in SPE, highlighting a trade-off between capturing more positives and maintaining balanced classification.
  
  SPE results are presented in Fig. \ref{fig:lambda_gamma}(e)(f). On ADHD-200, SPE aligns with ACC and SEN, peaking at $\lambda = 0.002$ and $\gamma = 0.001$. ABIDE-I shows greater tolerance to changes in $\lambda$, maintaining competitive SPE across values. However, SPE decreases significantly when $\gamma > 0$, reaching the lowest point at $\gamma = 0.002$, which suggests that sparsity may disrupt the ability of HA-STA to correctly identify negative samples on this dataset.
  
  From these observations, it can be concluded that $\lambda = 0.002$ provides a solid baseline for both datasets. For ADHD-200, a small $\gamma$ value such as 0.001 is recommended, while for ABIDE-I, disabling $\gamma$ yields optimal performance. Overall, the results indicate that the impact of regularization is dataset-dependent, and appropriate tuning of $\lambda$ and $\gamma$ is crucial for achieving balanced diagnostic ACC, SEN, and SPE.

  \subsubsection{Effect of the number of hyperedges}
  
\begin{figure}[htbp]
      \centering
      \includegraphics[width=\columnwidth]{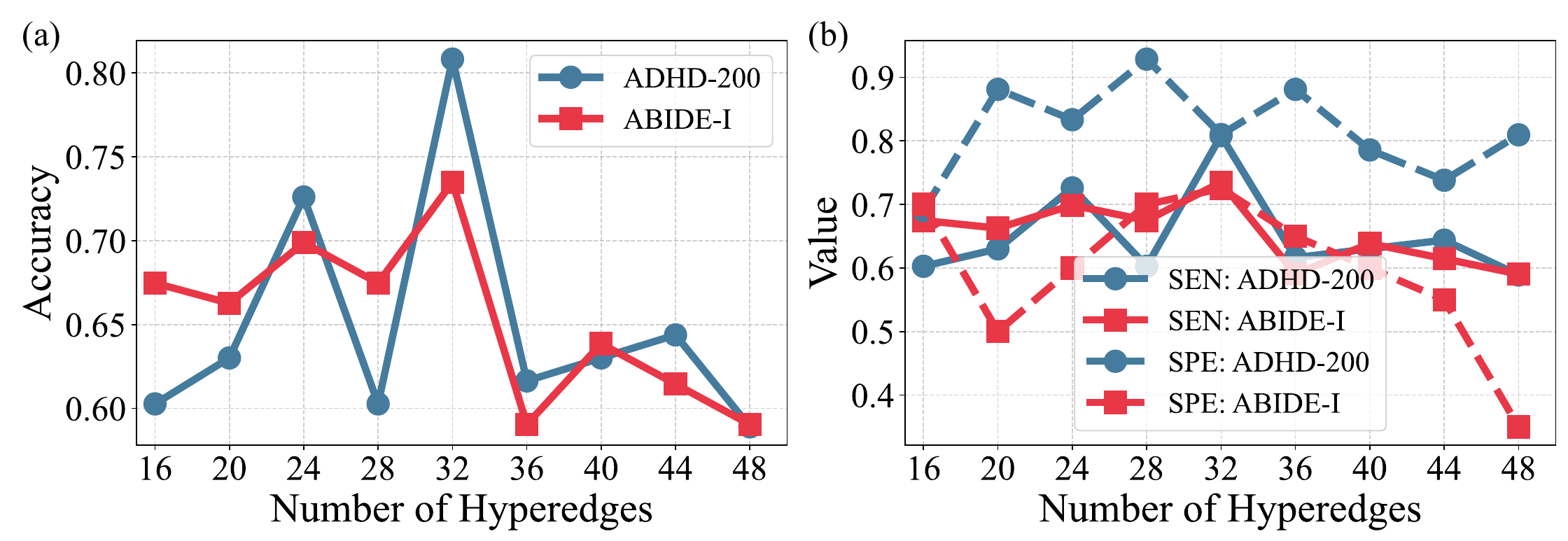}
      \caption{Compare the number of hyperedges of HA-STA in terms of ACC, SEN and SPE on ADHD-200 and ABIDE-I datasets.}
      \label{fig:K}
  \end{figure}

  We conduct a detailed analysis to explore the effect of the number of hyperedges on classification performance. As illustrated in Fig. \ref{fig:K}, we vary the number of hyperedges from 16 to 48 and evaluate the model on both the ADHD-200 and ABIDE-I datasets. The results reveal a consistent trend across both datasets: ACC, SEN, SPE improve steadily with the number of hyperedges up to a certain point, reaching a peak when the number of hyperedges is set to 32.
  This suggests that the optimal configuration not only enhances overall prediction correctness but also balances the ability of model to correctly identify both positive and negative cases.
  However, when the number of hyperedges exceeds 32, there is a notable decline across all three metrics. This performance drop suggests that an excessive number of hyperedges may introduce redundant or noisy connections, leading to overfitting or degraded representation quality.
  The consistent peak at 32 hyperedges across ACC, SEN, and SPE highlights the effectiveness and stability of the proposed model under this configuration, emphasizing the importance of carefully tuning this parameter to achieve optimal and reliable performance.
  
  \subsection{Hyperedge Degree Distribution Analysis}
  
\begin{figure}[htbp]
      \centering
      \includegraphics[width=\columnwidth]{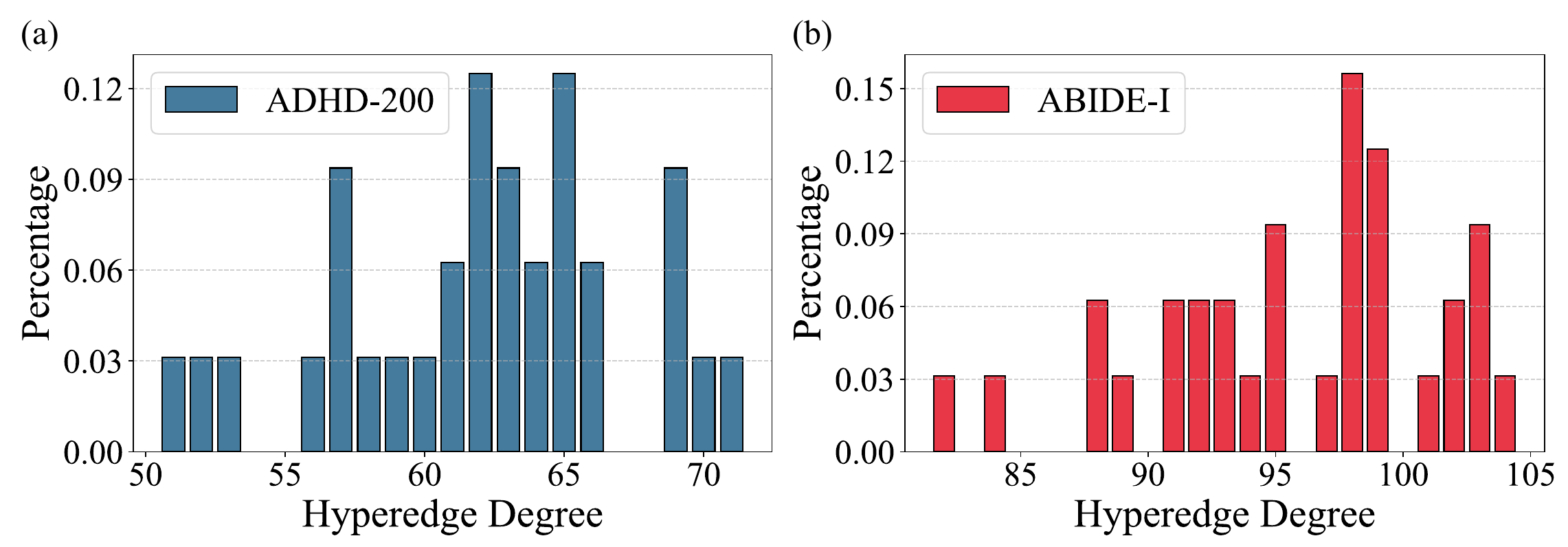}
      \caption{Hyperedge degree distribution of learned hyperedges on ADHD-200 and ABIDE-I datasets.}
      \label{fig:degree}
  \end{figure}

  Fig. \ref{fig:degree} illustrates the distribution of hyperedge degrees for both datasets. In the ADHD-200 dataset, most hyperedges connect between 55 and 70 brain regions, with a clear peak around 62 nodes. In the ABIDE dataset, hyperedges typically connect between 85 and 105 regions, peaking near 95. The relatively large number of connected nodes per hyperedge suggests that the model effectively captures global patterns of functional connectivity. However, connecting too many nodes may also introduce irrelevant or noisy signals, which could reduce discriminative power. The sparsity regularization term applied to hyperedge degrees addresses this issue by encouraging the model to avoid excessively large hyperedges (see Section \ref{regularization_hyperparameters} for sparsity regularization analysis). This helps ensure that each hyperedge remains both informative and focused, promoting a balance between global integration and local specificity—an important factor in distinguishing between healthy and diseased subjects.

  \subsection{Most Discriminative ROIs Analysis}
\begin{figure}[htbp]
    \centering
    \includegraphics[width=\columnwidth]{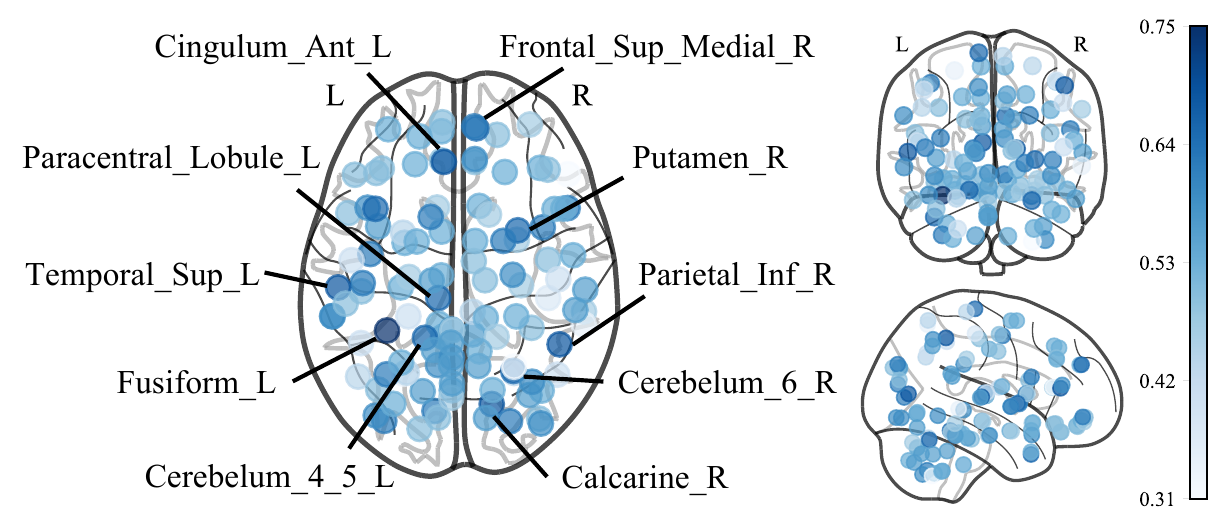}
    \caption{Visualize the regional importance of each region under the diagnosis of ADHD. The top ten most discriminative nodes are labeled in figure.}
    \label{fig:node}
  \end{figure}

  We used the ADHD-200 dataset in the physiological analysis because the adopted AAL-116 template is based on anatomical divisions, where each region corresponds closely to known functions and structures, contributing to the physiological interpretation of the results.
  By analyzing the regional importance of the 32 learned hyperedges, we identified 10 regions with the highest occurrence rates as shown in Fig. \ref{fig:node}, indicating their critical discriminative power for ADHD classification. 
  A deeper neurobiological analysis reveals that these regions cluster into several functional systems relevant to ADHD pathophysiology:
  
  
  
  \begin{itemize}
    \item \textbf{Executive Dysfunction and Salience Processing:}
    \textit{Frontal\_Sup\_Medial\_R} and \textit{Cingulum\_Ant\_L} are central to executive control and conflict monitoring. Structural and functional abnormalities in these regions have been consistently linked to attentional dysregulation in ADHD \cite{bush2005functional}. \textit{Putamen\_R}, involved in reward and motor regulation, shows altered activity associated with impulsivity \cite{bush2005functional}.
    
    \item \textbf{Sensorimotor and Attention Modulation:}
    \textit{Paracentral\_Lobule\_L} and \textit{Parietal\_Inf\_R} are implicated in motor planning and attention shifting. These areas exhibit atypical activation during cognitive flexibility tasks in ADHD, reflecting deficits in adaptive control \cite{hart2013meta}.
    
    \item \textbf{Sensory and Perceptual Processing:}
    \textit{Temporal\_Sup\_L}, \textit{Fusiform\_L}, and \textit{Calcarine\_R} are involved in auditory and visual processing. Reduced activity in these regions has been linked to impairments in sensory integration and visual attention in ADHD \cite{hale2014visual, wang2021rich}.
    
    \item \textbf{Cerebellar Timing and Regulation:}
    \textit{Cerebelum\_6\_R} and \textit{Cerebelum\_4\_5\_L} support motor and cognitive timing. Cerebellar abnormalities, are well documented in ADHD and relate to deficits in behavioral regulation \cite{valera2007meta}.
  \end{itemize}

  
  \subsection{Most Discriminative Hyperedge Analysis}
  
  The most discriminative hyperedge, identified by ranking the hyperedge importance, reflects a structured pattern of inter-regional connectivity. Rather than emphasizing isolated regional abnormalities, this hyperedge reflects deficits in multiple regions as a whole (this pattern is visually illustrated in Fig. \ref{fig:hyperedge}), highlighting a complex interplay among neural systems involved in attention regulation, executive control, emotion processing, sensory filtering, and motor coordination—domains frequently impaired in ADHD \cite{rubia2018cognitive}:
  
  \begin{figure*}[htbp]
      \centering
      \includegraphics[width=\textwidth]{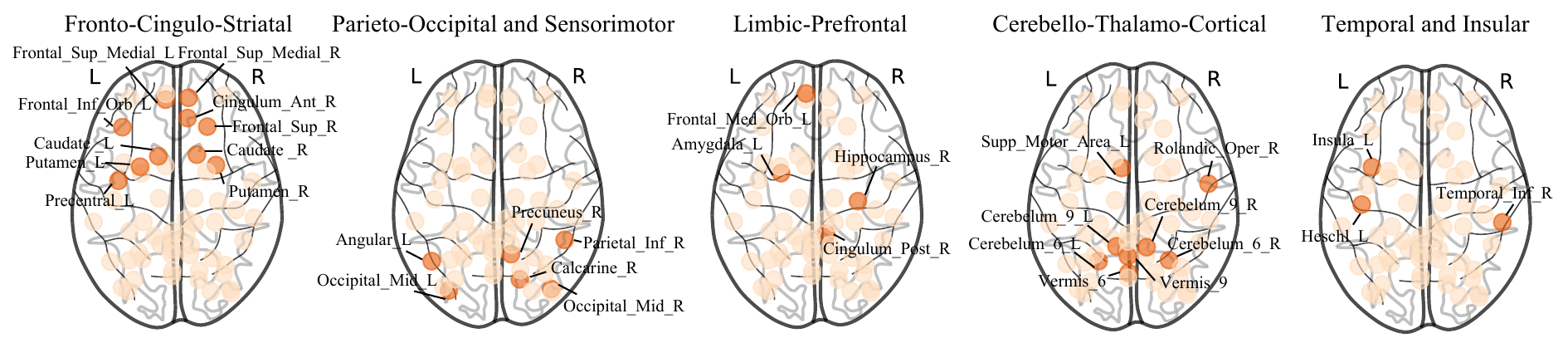}
      \caption{Visualization of the most discriminative hyperedge for ADHD diagnosis. All nodes in each graph form this most discriminative hyperedge. Brightly colored nodes indicate inter-regional connectivity associated with ADHD pathology.}
      \label{fig:hyperedge}
  \end{figure*} 
  
  \textbf{1. Fronto-Cingulo-Striatal Circuitry:} 
  The hyperedge includes a constellation of frontal regions (e.g., \textit{Precentral\_L}, \textit{Frontal\_Sup\_R}, \textit{Frontal\_Sup\_Medial\_L/R}, \textit{Frontal\_Inf\_Orb\_L}), anterior cingulate (\textit{Cingulum\_Ant\_R}), and basal ganglia components (\textit{Putamen\_L/R}, \textit{Caudate\_L/R}). These areas collectively form a canonical cognitive control network responsible for conflict monitoring, response inhibition, and executive regulation. The inclusion of both medial and orbital prefrontal cortices, in conjunction with striatal nodes, aligns with findings that ADHD is associated with underactivation in fronto-striatal loops and deficient salience detection \cite{vaidya1998selective, rubia1999hypofrontality}. 
  
  \textbf{2. Parieto-Occipital and Sensorimotor Integration:} 
  Regions such as \textit{Parietal\_Inf\_R}, \textit{Angular\_L}, \textit{Precuneus\_R}, \textit{Calcarine\_R}, and \textit{Occipital\_Mid\_L/R} appear jointly within the hyperedge, indicating disrupted integration between dorsal attention systems and visual-sensory regions. The \textit{Precuneus\_R}, a Default Mode Network (DMN) hub, is notably implicated in internally directed thought, and its engagement here may reflect abnormal task-DMN interaction. This is consistent with evidence showing that individuals with ADHD fail to appropriately deactivate the DMN during goal-directed tasks, leading to attention lapses and reduced cognitive efficiency \cite{fassbender2009lack, christakou2013disorder}.
  
  \textbf{3. Limbic-Prefrontal Emotional Regulation Network:} 
  The co-occurrence of \textit{Amygdala\_L}, \textit{Hippocampus\_R}, \textit{Cingulum\_Post\_R}, and \textit{Frontal\_Med\_Orb\_L} within the same hyperedge suggests impaired emotion-cognition integration. These regions are associated with affective salience, memory contextualization, and reward processing. Prior research has shown that abnormalities in limbic-prefrontal connectivity may underlie emotional impulsivity and poor motivational control in ADHD \cite{spencer2017abnormal}, contributing to behavioral dysregulation and mood instability.
  
  \textbf{4. Cerebello-Thalamo-Cortical Connectivity:} 
  The hyperedge robustly includes nodes from both cerebellar hemispheres (e.g., \textit{Cerebelum\_6\_L/R}, \textit{Cerebelum\_9\_L/R}, \textit{Vermis\_6}, \textit{Vermis\_9}) and cortical regions such as the \textit{Supp\_Motor\_Area\_L} and \textit{Rolandic\_Oper\_R}. These circuits contribute to predictive timing, sensorimotor coordination, and the modulation of cognitive effort. Consistent with cerebellar theories of ADHD, disruptions in cerebello-thalamo-cortical loops are thought to impair both motor and cognitive timing mechanisms in affected individuals \cite{rubia2014imaging}.
  
  \textbf{5. Temporal and Insular Contributions:} 
  The presence of \textit{Temporal\_Inf\_R}, \textit{Heschl\_L}, and \textit{Insula\_L} points to atypical sensory processing and interoceptive integration. The insula, as a key hub in the salience network, modulates the switch between the DMN and task-positive networks. Aberrant insular connectivity in ADHD may therefore contribute to inefficient network switching and diminished attentional control, corroborating prior findings of disrupted salience attribution in ADHD populations \cite{cortese2012toward, hart2012meta}.
  


  \section{Conclusion}
  In this paper, we propose a novel hypergraph attention-based spatio-temporal aggregation framework to jointly learn sparse and informative high-order brain structures from fMRI data. Guided by a MIMR objective, our method captures disease-relevant patterns while reducing redundancy. Experimental results on benchmark datasets demonstrate superior performance over the state-of-the-art approaches. The discovered spatio-temporal hyperedges are meaningful for neurological disorder analysis and offer insights into brain network organization. Our framework also holds promise for extension to multi-template and multimodal neuroimaging tasks.

  \bibliographystyle{IEEEtran}
  \bibliography{reference.bib}
  
  \end{document}